\documentclass[12pt]{article}
\usepackage{graphicx}
\usepackage{dcolumn}
\usepackage{bm}
\usepackage{amssymb}
\usepackage{amsmath}
\newcommand{\bq}{\begin{equation}}
\newcommand{\eq}{\end{equation}}

\newcommand{\bqr}{\begin{eqnarray}}
\newcommand{\eqr}{\end{eqnarray}}

\newcommand{\bqrx}{\begin{eqnarray*}}
\newcommand{\eqrx}{\end{eqnarray*}}

\newcommand{\br}{\begin{array}}
\newcommand{\er}{\end{array}}

\newcommand{\voo}{\vspace*{5pt}}

\begin{document}

\pagestyle{empty}

\setlength{\parindent}{18pt}
\setlength{\footskip}{.5in}



\vspace*{.6in}
\begin{center}
On limiting distributions of quantum Markov chains
\end{center}
\begin{center}
Chaobin\, Liu and \,Nelson Petulante\\
Department of Mathematics\\
Bowie State University\\
14000 Jericho Park Road\\
 Bowie, MD 20715 USA\\ 
{\footnotesize \,\,cliu@bowiestate.edu, npetulante@bowiestate.edu}
\voo\\
\voo\voo\voo
\end{center}
ABSTRACT:\ \ In a quantum Markov chain, the temporal succession of states is modeled by the repeated action of a ``bistochastic quantum operation" on the density matrix of a quantum system. Based on this conceptual framework, we derive some new results concerning the evolution of a quantum system, including its long-term behavior. Among our findings is the fact that the Ces$\grave{a}$ro limit of any quantum Markov chain always exists and equals the orthogonal projection of the initial state upon the eigenspace of the unit eigenvalue of the bistochastic quantum operation. Moreover, if the unit eigenvalue is the only eigenvalue on the unit circle, then the quantum Markov chain converges in the conventional sense to the said orthogonal projection. As a corollary, we offer a new derivation of the classic result describing limiting distributions of unitary quantum walks on finite graphs (Aharonov et al., 2001 \cite{AAKV01}).



\section{Introduction}        
The theory of Markov chains, when appropriately generalized, provides a potent paradigm for analyzing the stochastic evolution of quantum systems. Over the past decade, motivated largely by the prospect of super-efficient algorithms, the theory of so-called {\em quantum Markov chains}, especially in the guise of quantum walks,  has generated a huge volume of research, including many discoveries of fundamental importance, such as \cite{AAKV01,ABNVW01,NV00,FG98,CFG02,K03,A03,K06,VA08,K08} and numerous recent advances, such as \cite{MG10,SK10,SBC10,AAAd09,LP2009}. 

In the context of quantum walks, the itinerary of the walker is confined to a particular topological network. The walker's every move, from node to adjacent node, is governed by a set of local rules. When applied repeatedly to a given initial state of the system (represented by a superposition of basis states), these rules yield a succession of new states, reflecting, ad infinitum, the evolution of the system. A transition rule can be either unitary (closed) or non-unitary (open), depending, respectively, on whether it is intrinsic to the system or exposes the system to external influences such as decoherence, noise or measurement. 

In this paper, we adopt the formalism of ``quantum operations", whereby both unitary and non-unitary rules of state transition, as well as various combinations thereof, are treated under a unified mathematical model. In this framework, the ``transition matrix" of a classical Markov chain is replaced by a ``bistochastic quantum operation" and the ``state distribution vector" of the classical Markov chain is replaced by a ``density matrix".  The resulting description of quantum state evolution, known as a {\em quantum Markov chain} \cite{AAKV01,AA05}, turns out to resemble very closely the evolution of a classical Markov chain. In particular, the same kind of questions pertain, including questions about the long-term evolution of the process and the possible existence of a limiting state.

Among our findings is the fact that the Ces$\grave{a}$ro limit of any quantum Markov chain converges always to a stationary ``state", regardless of the initial state. As a noteworthy special case of this result, we remark that for any unitary quantum walk on a graph, as in \cite{AAKV01}, the limit of the time-averaged probability distribution always exists. 

To complete the picture, we specify conditions for the existence of a limiting state in the strict (non-Ces$\grave{a}$ro) sense of the word ``limiting". In the strict sense, it turns out that the limiting behavior depends only on the deployment on the unit disc of the eigenvalues of the bistochastic quantum operation. Specifically, if $\lambda=1$ is the only eigenvalue on the unit circle, then, for any given initial state, the associated quantum Markov chain converges to a stationary state. Moreover, if the eigenspace of $\lambda=1$ is one-dimensional or contains only a single density operator, then the associated Markov chain converges to the maximally mixed state, irrespective of the initial state. Otherwise, if the bistochastic quantum operation possesses any eigenvalue on the unit circle other than $\lambda=1$, then a limiting state is not guaranteed to exist except in the generalized sense of Ces$\grave{a}$ro. These findings are seen to be analogous, in a very natural way, to the fundamental properties of classical Markov chains (e.g. in \cite{CM00}, Chapter 8).

Our results may represent substantial progress toward answering the first of a set of ``open questions" posed by Ambainis in 2005 \cite{AA05}. We would be remiss not to acknowledge our indebtedness to \cite{NAJ10}, which treats the special case of a quantum Markov chain generated by a {\em random unitary operation}.

In what follows, after a brief review of some preliminaries (Section 2), we proceed (Section 3) to present our main results on the question of the limiting behavior of a quantum Markov chain. The proofs of the results given in Section 3 are deferred to Section 4. In Section 5, we present a comprehensive classification of quantum operations according to their limiting behavior. Finally, in Section 6, we offer some concluding remarks, including some relevant questions for further investigation.

\section{Bistochastic Quantum operations}

Given a Hilbert space $\mathcal{H}$ of finite dimension $N$, let $\mathfrak{B}(\mathcal{H})$ denote the set of all linear operators on $\mathcal{H}$, with inner product defined by
\begin{eqnarray}
\langle X,Y \rangle \equiv \mathrm{tr}( X^{\dagger}Y).\label{innerproduct}
\end{eqnarray}

The corresponding norm, called {\em Frobenius norm} or {\em Schatten} 2-{\em norm}, is defined by  
\begin{eqnarray}
\|X\| \equiv [\mathrm{tr}( X^{\dagger}X)]^{1/2}=\sqrt{\langle X,X \rangle}.
\end{eqnarray}
This choice of norm on $\mathfrak{B}(\mathcal{H})$ will remain in effect throughout this paper. 

Let $\mathfrak{D}(\mathcal{H}) \subset \mathfrak{B}(\mathcal{H})$ denote the set of {\em positive} operators $\rho: \mathcal{H}\rightarrow \mathcal{H}$ with Tr$(\rho)=1$. The operators $\rho \in \mathfrak{D}(\mathcal{H})$ are the so-called ``density operators". They serve to model, as faithfully as do the ``state vectors" themselves, the possible states of a quantum system whose state vectors reside in $\mathcal{H}$.  
 
By a {\em super-operator} ${\bf \Phi}$ on $\mathfrak{B}(\mathcal{H})$, we mean a linear mapping ${\bf \Phi}: \mathfrak{B}(\mathcal{H})\rightarrow\mathfrak{B}(\mathcal{H})$, with norm defined by  
\begin{eqnarray}
\|{\bf \Phi}\| \equiv \mathrm{Sup}_{X\in \mathfrak{B}(\mathcal{H})}\frac{\|{\bf \Phi}(X)\|}{\|X\|}.
\end{eqnarray}

Note that $\mbox{dim}{\mathfrak{B}(\mathcal{H})}=N^{2}$, where $N=\mbox{dim}(\mathcal{H})$. Thus, any super-operator ${\bf \Phi}$ on $\mathfrak{B}(\mathcal{H})$ can be represented, relative to a given basis for $\mathfrak{B}(\mathcal{H})$, by an $N^{2}\times N^{2}$ matrix. In the sequel, this matrix will be denoted by the symbol $\left[{\bf \Phi}\right]$. In particular, relative to a special basis consisting of eigenvectors and generalized eigenvectors of ${\bf \Phi}$, the shape of the matrix $\left[{\bf \Phi}\right]$ conforms to a special quasi-diagonal lay-out called the Jordan canonical form. The details can be found in any one of a number of sources, including \cite{NAJ10}.  

Among the set of super-operators, we distinguish a special subset called ``quantum operations". By definition, to qualify as a quantum operation, the super-operator ${\bf \Phi}$ must be {\em completely positive}, meaning that the extended map ${\bf \Phi}\otimes \mathbb{I}_n$ is positive for all $n\geq 1$. 

The formalism of quantum operations is flexible enough to handle both unitary (closed) and non-unitary (open), or a mixture thereof, of discrete transitions of state of a quantum system. For a good introductory exposition of this subject, see \cite{NC00,P08}.

By Choi's Theorem \cite{Choi1975} and \cite{NC00,P08,K1983}, any completely positive linear operator, including any quantum operation ${\bf \Phi}: \mathfrak{B}(\mathcal{H})\rightarrow\mathfrak{B}(\mathcal{H})$, can be represented in terms of a set $\mathcal{A}=\{A_i \,|\,\, i=1,2,..., N^{2}\}$ of ``Kraus operators", as follows:  

\begin{eqnarray}
{\bf \Phi}_{\mathcal{A}}(X)=\sum_{i}A_iX A_i^{\dagger}.\label{choi-ex}
\end{eqnarray}
In this expression, which we call the ``Choi expansion" of ${\bf \Phi}$, the symbol $A_i^{\dagger}$ denotes ${\bar{A_i}}^{T}$(transpose of the complex conjugate of $A_i$). 
 
In terms of the Choi expansion, the condition of being {\em trace-preserving}, meaning that $\mbox{Tr}({\bf \Phi}_{\mathcal{A}}(X))=\mbox{Tr}(X)$ for all $X \in \mathfrak{B}(\mathcal{H})$, is equivalent to the condition:

\begin{equation}
{\sum_{i}A_i^{\dagger}A_i=\mathbb{I}_N}. 
\end{equation}

On the other hand, if the Kraus operators of ${\bf \Phi}_{\mathcal{A}}$ satisfy the dual condition:

\begin{equation}
{\sum_{i}A_i A_i^{\dagger}=\mathbb{I}_N}, 
\end{equation} 
then ${\bf \Phi}_{\mathcal{A}}$ is said to be {\em unital}. Note that (6) is equivalent to the simple statement that ${\bf \Phi}_{\mathcal{A}}(\mathbb{I}_N)=\mathbb{I}_N$. 

A quantum operation which is both unital and trace-preserving is called {\em bistochastic}. The term {\em doubly stochastic quantum channel} also has been used to refer to quantum operations of this sort \cite{JW09}. Note that a bistochastic quantum operation transforms elements of $\mathfrak{D}(\mathcal{H})$ into elements of $\mathfrak{D}(\mathcal{H})$. In other words, since the elements of $\mathfrak{D}(\mathcal{H})$ represent the states of a quantum system, a bistochastic quantum operation transforms states of a quantum system into other legitimate states of that system.  

\section{Limit theorems for bistochastic quantum operations}


In the sequel, the proofs of all theorems, corollaries and supporting lemmas are deferred to Section 4. 

By \cite{PGWPR06}, a bistochastic quantum operation ${\bf \Phi}_{\mathcal{A}}$ must satisfy the condition $\|{\bf \Phi}_{\mathcal{A}}\|\leq 1$. Thus the spectrum of ${\bf \Phi}_{\mathcal{A}}$ is confined to the unit disk. Also, since ${\bf \Phi}_{\mathcal{A}}(\mathbb{I}_N)=\mathbb{I}_N$, we see that $\lambda = 1$ is an eigenvalue of ${\bf \Phi}_{\mathcal{A}}$ and $\|{\bf \Phi}_{\mathcal{A}}\|=1$. For future reference, we record these observations in the form of a lemma:
\vskip 0.1in

{\bf Lemma~1.} Let ${\bf \Phi}_{\mathcal{A}}$ be a bistochastic quantum operation on the Hilbert space $\mathfrak{B}(\mathcal{H})$, then
\begin{enumerate}
  \item  $\|{\bf \Phi}_{\mathcal{A}}\|=1$
  \item If $\lambda$ is an eigenvalue of ${\bf \Phi}_{\mathcal{A}}$, then $|\lambda|\le 1$.
  \item The value $\lambda =1$ is an eigenvalue of ${\bf \Phi}_{\mathcal{A}}$.
\end{enumerate}
\vskip 0.1in

The observations recorded in Lemma 1 are new by no means. See, for instance, \cite{BCSZ09}.



For an eigenvalue $\lambda$ of ${\bf \Phi}_{\mathcal{A}}$, let $\mathsf{Ker}({\bf \Phi}_{\mathcal{A}}-\lambda\mathbb{I})$ and $\mathsf{Ran}({\bf \Phi}_{\mathcal{A}}-\lambda\mathbb{I})$ denote, respectively, the kernel and range of the operator ${\bf \Phi}_{\mathcal{A}}-\lambda\mathbb{I}$ on the Hilbert space $\mathfrak{B}(\mathcal{H})$.
\vskip 0.1in

{\bf Lemma~2.}\,\, Let ${\bf \Phi}_{\mathcal{A}}$ be a bistochastic quantum operation on $\mathfrak{B}(\mathcal{H})$ and let $\lambda$ be an eigenvalue of ${\bf \Phi}_{\mathcal{A}}$ with $|\lambda|=1$. Then $\mathsf{Ker}({\bf \Phi}_{\mathcal{A}}-\lambda\mathbb{I})\cap\mathsf{Ran}({\bf \Phi}_{\mathcal{A}}-\lambda\mathbb{I})=\{0\}$. 
\vskip 0.1in

From the preceding lemma, we can derive an important inference concerning the algebraic and geometric multiplicities of the eigenvalues of ${\bf \Phi}_{\mathcal{A}}$. For an eigenvalue $\lambda$ of ${\bf \Phi}_{\mathcal{A}}$, let its algebraic multiplicity be denoted by $m(\lambda)$ and lets its geometric multiplicity be denoted by $g(\lambda)$. Recall that $g(\lambda)=\mathrm{dim}\mathsf{Ker}({\bf \Phi}_{\mathcal{A}}-\lambda\mathbb{I})$. In addition, let the spectrum of ${\bf \Phi}_{\mathcal{A}}$ be denoted by $\Lambda({\bf \Phi}_{\mathcal{A}})$ and let $\Lambda_{1}({\bf \Phi}_{\mathcal{A}})$ denote the subset of $\Lambda({\bf \Phi}_{\mathcal{A}})$ consisting of $\lambda \in \Lambda({\bf \Phi}_{\mathcal{A}})$ with $|\lambda|=1$.
\vskip 0.1in

{\bf Lemma~3.}\,\, If $\lambda\in \Lambda_{1}({\bf \Phi}_{\mathcal{A}})$, where ${\bf \Phi}_{\mathcal{A}}$ is a bistochastic quantum operation on $\mathfrak{B}(\mathcal{H})$, then $m(\lambda)=g(\lambda)$.
\vskip 0.1in

As in the proof of Lemma 3, let $\left[{\bf \Phi}_{\mathcal{A}}\right]$ denote the $N^{2}\times N^{2}$ Jordan canonical matrix representation of ${\bf \Phi}_{\mathcal{A}}$. To conserve type-set space, and without undue risk of confusion to the reader, we prefer to display the matrix $\left[{\bf \Phi}_{\mathcal{A}}\right]$ in the following self-explanatory format:
\begin{eqnarray}
\left[{\bf \Phi}_{\mathcal{A}}\right]=\mbox{diag}\left(\lambda_1, \lambda_2, ..., \lambda_k, J_1, J_2, ...,J_h\right), 
\label{JCF}\end{eqnarray}
where $\lambda_i\in \Lambda_{1}({\bf \Phi}_{\mathcal{A}})$ and $J_{r}$, $r=1,2,...,h$, denote the Jordan blocks corresponding to eigenvalues whose norms are strictly less than unity.

For an eigenvalue $\lambda$ of ${\bf \Phi}_{\mathcal{A}}$, let $E_{{\bf \Phi}_{\mathcal{A}}}(\lambda)=\mathsf{Ker}({\bf \Phi}_{\mathcal{A}}-\lambda \mathbb{I})$ denote the eigenspace of $\lambda$. In particular, for $\lambda=1$, it has been established \cite{AGG02,Kribs03,HKL04} that 
\begin{eqnarray}
E_{{\bf \Phi}_{\mathcal{A}}}(1)=\mathsf{Ker}({\bf \Phi}_{\mathcal{A}}-\mathbb{I})=\{X\in \mathfrak{B}(\mathcal{H}): XA_i=A_iX\, ; i=1,2,...\}. \label{eigenspaceof1}
\end{eqnarray}

The following lemma articulates the special status of $E_{{\bf \Phi}_{\mathcal{A}}}(1)$ relative to the other eigenspaces of ${\bf \Phi}_{\mathcal{A}}$.\vskip 0.1in

{\bf Lemma~4.}\, Let $\lambda$ be an eigenvalue of ${\bf \Phi}_{\mathcal{A}}$ such that $|\lambda|=1$ and $\lambda \ne 1$. Let $\alpha$ be an eigenvalue of ${\bf \Phi}_{\mathcal{A}}$ with $|\alpha|<1$. Let $Y_{1}$, ..., $Y_{j_{\alpha}}$ denote the generalized eigenvectors belonging to $\alpha$. Then 
\begin{enumerate}
  \item $E_{{\bf \Phi}_{\mathcal{A}}}(1)\perp E_{{\bf \Phi}_{\mathcal{A}}}(\lambda)$
  \item $E_{{\bf \Phi}_{\mathcal{A}}}(1)\perp \mathsf{Span}\{Y_{1}, ..., Y_{j_{\alpha}}\}$
\end{enumerate}
\vskip 0.1in

In other words, by Lemma 4, the eigenspace $E_{{\bf \Phi}_{\mathcal{A}}}(1)$ is orthogonal to every eigenvector (including every generalized eigenvector) of ${\bf \Phi}_{\mathcal{A}}$ belonging to eigenvalues other than $\lambda =1$. 
\vskip 0.1in


{\bf Theorem~5.}\,\, Let ${\bf \Phi}_{\mathcal{A}}$ be a bistochastic quantum operation on the Hilbert space $\mathfrak{B}(\mathcal{H})$ and let $\rho(0)\in \mathfrak{D}(\mathcal{H})$ denote the density matrix representing the initial state of a quantum system. Let $g(1)$ denote the geometric multiplicity of the eigenvalue $\lambda=1$ and let  $\{Z_{r}\}_{1\leq r\leq g(1)}$ denote an orthonormal basis for $E_{{\bf \Phi}_{\mathcal{A}}}(1)$. If $\lambda =1$ is the only eigenvalue of ${\bf \Phi}_{\mathcal{A}}$ on the unit circle, then the iterated succession of quantum states $\rho(t)={\bf \Phi}_{\mathcal{A}}^t\rho(0)$ converges to $\rho(\infty)=\sum_{l=1}^{g(1)}\mathrm{tr}( Z_l^{\dagger}\rho(0))Z_l$. In particular, if $g(1)=1$, then $\lim_{t\rightarrow \infty}{\bf \Phi}_{\mathcal{A}}^t\rho(0)=\frac{1}{N}\mathbb{I}$, independently of the initial state $\rho(0)$.
\vskip 0.1in


In previous publications, such as \cite{SPWC10}, the special case of a bistochastic quantum operation whereby $g(1)=1$ is known as a {\it primitive quantum channel}. The reference \cite{SPWC10} provides an excellent analysis of this line of investigation.    

In the literature, numerous examples can be found of bistochastic quantum operations (a.k.a. quantum channels) for which the limiting behavior of the associated quantum Markov chain is governed by Theorem 5. Among these examples are 1) the {\it quantum walk} on a cycle, subject to decoherence on the degree freedom of coin \cite {LP2010E} and 2) the generalized  {\it depolarizing channel} on a multiple-dimensional quantum system \cite{NC00,P08}. These examples can be treated as special cases of the following corollary to Theorem 5. 
\vskip 0.1in

{\bf Corollary~6.}\,\, If the bistochastic quantum operation ${\bf \Phi}_{\mathcal{A}}$ is given by 

\begin{eqnarray}
{\bf \Phi}_{\mathcal{A}}(X)=(1-p)X+\sum_{i}A_iX A_i^{\dagger},
\end{eqnarray}

where $\sum_{i}A_iA_i^{\dagger}=p\mathbb{I}$ and $0<p<1$, then, as in Theorem 5,  $\rho(\infty)=\sum_{r=1}^{g(1)}\mathrm{tr}( Z_r^{\dagger}\rho(0))Z_r$. 
\vskip 0.1in

In the context of quantum channels, the Kraus operators defining the generalized {\it depolarizing channel} can be scalar multiples of the discrete Weyl operators (or the generalized Pauli operators). When such is the case, as in for instance \cite{JW09}, the eigenspace of the eigenvalue 1 must be one-dimensional. Consequently the associated quantum Markov chain must converge to the maximally mixed state, regardless of the initial state.




\vskip 0.1in



The main assertion of Theorem 5 pertains only to bistochastic quantum operations ${\bf \Phi}_{\mathcal{A}}$ possessing no eigenvalues on the unit circle other than $\lambda=1$. In general, when this condition is relaxed to admit other eigenvalues on the unit circle, the expression  $\rho(t)={\bf \Phi}_{\mathcal{A}}^t\rho(0)$ is compelled no longer to converge to a stationary state. Many examples can be found in the literature, including classic cases of unitary quantum walks on finite graphs \cite{AAKV01}. 

However, even if a limiting state fails to exist in the usual sense, we still might want to probe the possibility of a ``limiting state" $\tilde{\rho}(\infty)$ in the sense of Ces$\grave{a}$ro:
\begin{eqnarray}
\tilde{\rho}(\infty)=\lim_{t\rightarrow \infty}\frac{{\bf \Phi}_{\mathcal{A}}\rho(0)+{\bf \Phi}_{\mathcal{A}}^2\rho(0)+...+{\bf \Phi}_{\mathcal{A}}^t\rho(0)}{t}. \label{cesaro}
\end{eqnarray}

In terms of this generalized sense of ``limiting state", it turns out that every quantum Markov chain converges.   
\vskip 0.1in

{\bf Theorem~7.}\,\, Let ${\bf \Phi}_{\mathcal{A}}$ be a bistochastic quantum operation on the Hilbert space $\mathfrak{B}(\mathcal{H})$ and let $\rho(0)\in \mathfrak{D}(\mathcal{H})$ denote the density matrix representing the initial state of a quantum system. Let $g(1)$ denote the geometric multiplicity of the eigenvalue $\lambda=1$ and let  $\{Z_{r}\}_{1\leq r\leq g(1)}$ denote an orthonormal basis for $E_{{\bf \Phi}_{\mathcal{A}}}(1)$. Then, for every $\rho(0)\in \mathfrak{D}(\mathcal{H})$: 
\begin{eqnarray}
\tilde{\rho}(\infty)=\lim_{t\rightarrow \infty}\frac{1}{t}\sum_{n=1}^{t}{\bf \Phi}_{\mathcal{A}}^n\rho(0)=\sum_{l=1}^{g(1)}\mathrm{tr}( Z_l^{\dagger}\rho(0))Z_l.
\end{eqnarray}
\vskip 0.1in

According to this theorem, the sequence of Ces$\grave{a}$ro means $\{\frac{1}{t}\sum_{n=1}^{t}{\bf \Phi}_{\mathcal{A}}^n\}_{t=1}^{\infty}$ is guaranteed always to converge to a limit which coincides with the orthogonal projection of the initial state upon the eigenspace of the unit eigenvalue of the bistochastic quantum operation. It is only fair to point out that alternate versions of this result have appeared in the literature. For instance, in \cite{AGG02} (see Theorem 2.4(a)) it is shown that there exists a subsequence of the full sequence of Ces$\grave{a}$ro means which converges to a limit within the space of fixed points of the quantum operation. A closer match to Theorem 7, derived by different means and framed in different language, can be found in Chapter 6 of \cite{MW10}.

\vskip 0.1in

As an immediate corollary of Theorem 7, it follows, as in \cite{AAKV01}, that the time-averaged probability distribution for any unitary quantum walk on a finite graph must converge. For a unitary quantum walk on a finite graph starting with an initial state $|\alpha_0\rangle$, let $P_n(v|\alpha_0)$ denote the probability of finding the walker at node $v$ at time $n$, and let $\bar{P}_t(v|\alpha_0)=\frac{1}{t}\sum_{n=1}^{t}P_n(v|\alpha_0)$ denote the time-averaged probability. 
\vskip 0.1in

{\bf Corollary~8.}\, (see Theorem 3.4., \cite{AAKV01})\, Let $\psi_j$, $\lambda_j$ denote the unit eigenvectors and corresponding eigenvalues of the unitary operator $U$ associated with a unitary quantum walk on a finite graph and let the bistochastic quantum operation ${\bf \Phi}_{\mathcal{A}}$ be defined by ${\bf \Phi}_{\mathcal{A}}(\rho)=U\rho U^{\dagger}$. Then, for any initial state $|\alpha_0\rangle=\sum_ja_j|\psi_j\rangle$, we have
\begin{eqnarray}
\lim_{t\rightarrow \infty}\bar{P}_t(v|\alpha_0)=\sum_{i,j,a}a_ia^{\star}_j\langle a,v|\psi_i\rangle\langle \psi_j|a,v\rangle,
\end{eqnarray}
where the sum is restricted to pairs $i, j$ such that $\lambda_i=\lambda_j$.


\section{Proofs of the Theorems}

\noindent This section is reserved for the proofs of theorems, lemmas and corollaries given in Section 3.

\subsection{Proof of Lemma 2}
\noindent Suppose $X\in \mathsf{Ker}({\bf \Phi}_{\mathcal{A}}-\lambda\mathbb{I})\cap\mathsf{Ran}({\bf \Phi}_{\mathcal{A}}-\lambda\mathbb{I})$. Then ${\bf \Phi}_{\mathcal{A}}(X)=\lambda X$ and ${\bf \Phi}_{\mathcal{A}}(Y)-\lambda Y=X$ for some $Y\in \mathfrak{B}(\mathcal{H})$. Applying the linearity of ${\bf \Phi}_{\mathcal{A}}$, we infer that ${\bf \Phi}^n_{\mathcal{A}}(Y)=\lambda^{n}Y+n\lambda^{n-1}X$ for all $n\geq 1$. Consequently $\|n\lambda^{n-1}X\|=\|{\bf \Phi}^n_{\mathcal{A}}(Y)-\lambda^{n}Y\|$, which implies that $n\|X\|\le \|{\bf \Phi}_{\mathcal{A}}\|^n\|Y\|+\|Y\|$. But since $\|{\bf \Phi}_{\mathcal{A}}\|=1$ (see Lemma 1), we have $n\|X\|\leq 2\|Y\|$. Since this inequality must hold for all $n\geq 1$, we conclude that $\|X\|=0$, whence $X=0$. $\square$

In the above proof, we have borrowed liberally from the reasoning employed in \cite{NAJ10}, which treats the special case of a {\em random unitary operation}. 

\subsection{Proof of Lemma 3}
\noindent We proceed by contradiction. Suppose $m(\lambda)\neq g(\lambda)$. Let $\left[{\bf \Phi}_{\mathcal{A}}\right]$ denote the $N^{2}\times N^{2}$ matrix representation of ${\bf \Phi}_{\mathcal{A}}$ in Jordan canonical form. On the one hand, $\left[{\bf \Phi}_{\mathcal{A}}\right]$ must contain a Jordan block $J$ belonging to ${\lambda}$ of size $>1$. On the other hand, by standard matrix theory, there must exist a generalized eigenvector, say $v$, of ${\bf \Phi}_{\mathcal{A}}$ such that $({\bf \Phi}_{\mathcal{A}}-\lambda\mathbb{I})v$ is itself an eigenvector of ${\bf \Phi}_{\mathcal{A}}$. This implies that $\mathsf{Ker}({\bf \Phi}_{\mathcal{A}}-\lambda\mathbb{I})\cap\mathsf{Ran}({\bf \Phi}_{\mathcal{A}}-\lambda\mathbb{I})\ne\{0\}$, which contradicts Lemma 2. $\square$

\subsection{Proof of Lemma 4}

\noindent The adjoint operator of ${\bf \Phi}_{\mathcal{A}}$ is given by 
\begin{eqnarray}
{\bf \Phi}^{\dagger}_{\mathcal{A}}(X)=\sum_{i}A_i^{\dagger}X A_i.
\end{eqnarray}
Thus, by Eq. (\ref{eigenspaceof1}), we have $\mathsf{Ker}({\bf \Phi}_{\mathcal{A}}-\mathbb{I})=\mathsf{Ker}({\bf \Phi}^{\dagger}_{\mathcal{A}}-\mathbb{I})$.

To prove statement 1, let $Z \in \mathsf{Ker}({\bf \Phi}_{\mathcal{A}}-\mathbb{I})$ and $Y\in \mathsf{Ker}({\bf \Phi}_{\mathcal{A}}-\lambda\mathbb{I})$. Then $\langle Z, Y\rangle=\langle {\bf \Phi}^{\dagger}_{\mathcal{A}} Z, Y\rangle=\langle Z, {\bf \Phi}_{\mathcal{A}}Y\rangle=\lambda\langle Z,Y\rangle$. Since $\lambda \ne 1$, it follows that $\langle Z, Y\rangle=0$. Hence $\mathsf{Ker}({\bf \Phi}_{\mathcal{A}}-\mathbb{I})\perp \mathsf{Ker}({\bf \Phi}_{\mathcal{A}}-\lambda \mathbb{I})$.

We proceed to justify statement 2. For an eigenvalue $\alpha$ with $|\alpha|<1$, we may assume, without loss of generality, that the generalized eigenvectors belonging to $\alpha$ are arranged in a sequence $Y_1$, $Y_2$, ..., $Y_{j_{\alpha}}$ such that:
\begin{eqnarray}
({\bf \Phi}_{\mathcal{A}}-\alpha\mathbb{I})Y_r=Y_{r-1},
\end{eqnarray}
where, by definition, $Y_0 = 0$. It follows
that $({\bf \Phi}_{\mathcal{A}}-\alpha\mathbb{I})^r Y_r = 0$. If $Z\in \mathsf{Ker}({\bf \Phi}_{\mathcal{A}}-\mathbb{I})$, then $\langle Z, Y_1\rangle=\langle {\bf \Phi}_{\mathcal{A}}^{\dagger}Z, Y_1\rangle=\langle Z, {\bf \Phi}_{\mathcal{A}}Y_1\rangle=\alpha\langle Z, Y_1\rangle$, which implies that $\langle Z, Y_1\rangle=0$. Similarly, $\langle Z, Y_2\rangle=\langle {\bf \Phi}_{\mathcal{A}}^{\dagger}Z, Y_2\rangle=\langle Z, {\bf \Phi}_{\mathcal{A}}Y_2\rangle=\langle Z, Y_1\rangle+\alpha\langle Z, Y_2\rangle$, from which it follows that $\langle Z, Y_2\rangle=0$. Likewise, by the same reasoning, applied repeatedly, we deduce that $\langle Z, Y_r\rangle=0$ for any $r$. Thus $\mathsf{Ker}({\bf \Phi}_{\mathcal{A}}-\mathbb{I})\perp \mathsf{Span}\{Y_1, Y_2, ..., Y_{j_{\alpha}}\}$. $\square$


\subsection{Proof of Theorem 5}
\noindent Let $\alpha_1$, ..., $\alpha_h$ denote the eigenvalues of ${\bf \Phi}_{\mathcal{A}}$  whose absolute values are less than 1. Since, by Lemma 3, $\lambda=1$ is the only eigenvalue on the unit circle, a basis for $\mathfrak{B}(\mathcal{H})$ might be assembled as follows. Starting with an orthonormal basis for eigenspace $E_{{\bf \Phi}_{\mathcal{A}}}(1)$, append a maximal set of linearly independent generalized eigenvectors belonging to the eigenvalues $\alpha_r$, $r=1, 2, ..., h$. Then, relative to this basis, the Jordan canonical matrix representation of ${\bf \Phi}_{\mathcal{A}}$ is given by:
\begin{eqnarray}
\left[{\bf \Phi}_{\mathcal{A}}\right]=\mbox{diag}\left(\mathbb{I}_{g(1)}, J_1, J_2, ...,J_h\right), 
\end{eqnarray}
where $J_r$ is the Jordan block corresponding to the eigenvalue $\alpha_r$, $r=1, 2, ..., h$.

Consider what becomes of the Jordan blocks of the powers $\left[{\bf \Phi}_{\mathcal{A}}\right]^t$ as $t\rightarrow \infty$. Since each of the Jordan blocks $J_r$ is an upper triangular matrix whose diagonal is populated by a single eigenvalue of modulus strictly less than unity, it is a simple exercise in elementary algebra to show that $\lim_{t\rightarrow\infty}J^{t}_{r}=O_{r}$ (zero matrix of same size as $J_{r}$). Thus, if we define 
\begin{eqnarray}
\left[{\bf \Phi}^{\infty}_{\mathcal{A}}\right]=\mbox{diag}\left(\mathbb{I}_{g(1)}, O_1, O_2, ...,O_h\right), \label{asymptotic-form}
\end{eqnarray}
then $\|{\bf \Phi}^{t}_{\mathcal{A}}-{\bf \Phi}_{\mathcal{A}}^{\infty}\|\rightarrow 0$. 

Next, we consider the effect upon on the initial state $\rho(0)$ of ${\bf \Phi}^{t}_{\mathcal{A}}$ as $t\rightarrow\infty$. 


By Lemma 4, we may choose for $\mathfrak{B}(\mathcal{H})$ a basis consisting of a basis $Z_{1}, Z_{2}, ..Z_{g(1)}$ for $E_{{\bf \Phi}_{\mathcal{A}}}(1)$ together with an orthogonally complement basis consisting of generalized eigenvectors belonging to all other eigenvalues. In terms of such a basis we have $\rho(0)=\sum_{r=1}^{g(1)}c_r Z_r \oplus W$ where $W \perp Z_{r}$ for all $r=1, 2, ..., g(1)$.  By simple linear algebra, it follows that $c_r=\mathrm{tr}( Z_r^{\dagger}\rho(0))$ for $r=1, 2, ..., g(1)$. Therefore, by Eq. (\ref{asymptotic-form}), we have $\lim_{t\rightarrow \infty}{\bf \Phi}_{\mathcal{A}}^t\rho(0)=\sum_{r=1}^{g(1)}\mathrm{tr}( Z_r^{\dagger}\rho(0))Z_r$. 

Finally, if $\mathrm{dim}\mathsf{Ker}({\bf \Phi}_{\mathcal{A}}-\mathbb{I})=1$, then, to serve as a normalized basis for $E_{{\bf \Phi}_{\mathcal{A}}}(1)$, we may take $\{\frac{1}{\sqrt{N}}\mathbb{I}_N\}$, from which it follows that $\lim_{t\rightarrow \infty}{\bf \Phi}_{\mathcal{A}}^t\rho(0)=\frac{1}{N}\mathbb{I}$, for any initial state $\rho(0)$. $\square$


\subsection{Proof of Corollary 6}

\noindent It suffices to verify that $\lambda=1$ is the only eigenvalue of ${\bf \Phi}_{\mathcal{A}}$ on the unit circle. By \cite{PGWPR06}, whenever $\sum_{i}A_iA_i^{\dagger}=p\mathbb{I}$, then $\|\sum_{i}A_iXA_i^{\dagger}\|\le p\|X\|$ for any operator $X$. Now suppose $\lambda$ is an eigenvalue of ${\bf \Phi}_{\mathcal{A}}$ on the unit circle and $X$ is an eigenvector belonging to $\lambda$. Then ${\bf \Phi}_{\mathcal{A}}(X)=\lambda X=(1-p)X+\sum_{i}A_iX A_i^{\dagger}$, from which it follows that  $\|\lambda X\|=\|X\|=\|(1-p)X+\sum_{i}A_iX A_i^{\dagger}\|\le \|(1-p)X\|+\|\sum_{i}A_iX A_i^{\dagger}\|\le (1-p)\|X\|+p\|X\|=\|X\|$. Thus, all of the preceding inequalities actually are equalities. In particular, we have $\|\sum_{i}A_iX A_i^{\dagger}\|=p\|X\|$. Thus $(\lambda-1+p)X=\sum_{i}A_iX A_i^{\dagger}$, which implies that $|\lambda-1+p|\cdot\|X\|=p\|X\|$. It follows that $\lambda=1$. $\square$

\subsection{Proof of Theorem 7}
\noindent To evaluate the limit in Eq.(\ref{cesaro}), we have only to reexamine Eq.(\ref{JCF}). For every eigenvalue $\lambda=\lambda_{i}$ of ${\bf \Phi}_{\mathcal{A}}$, if $\lambda\ne 1$, then $\frac{1}{t}\sum_{n=1}^{t} \lambda^n=\frac{1}{t}\frac{\lambda-\lambda^{t+1}}{1-\lambda}\rightarrow 0$. Similarly, for every Jordan block $J=J_{r}$ in Eq.(\ref{JCF}), we have $\frac{1}{t}\sum_{n=1}^{t} J^n=\frac{1}{t}(\mathbb{I}-J)^{-1}(J-J^{t+1})\rightarrow 0$. Based on these two observations, we deduce that $\frac{1}{t}\sum_{n=1}^t [{\bf \Phi}_{\mathcal{A}}]^n$ converges to the diagonal matrix 
\begin{eqnarray}
{\bf \bar{\Phi}}_{\mathcal{A}}^{\infty} =
 \begin{bmatrix}
  \mathbb{I}_{g(1)} & 0 \\
  0& 0
 \end{bmatrix}.\label{casaro-form}
\end{eqnarray} 
The expression for  $\tilde{\rho}(\infty)$ follows by a path of reasoning identical to that employed in the proof of Theorem 5 to obtain the analogous expression for the limiting state in that situation. $\square$

\subsection{Proof of Corollary 8}

\noindent As an orthonormal basis for the eigenspace $E_{{\bf \Phi}_{\mathcal{A}}}(1)$, we may take the set of vectors $\bigcup_{i,j}\{|\psi_i\rangle \langle \psi_j|\}$, where the union is restricted to pairs $i, j$ such that $\lambda_i=\lambda_j$. Accordingly, the orthogonal projection of the initial state $\rho(0)=|\alpha_0\rangle\langle\alpha_0|$ into the eigenspace $E_{{\bf \Phi}_{\mathcal{A}}}(1)$ is 
\begin{eqnarray}
\sum_{i,j}\langle|\psi_i\rangle\langle \psi_j|,\rho(0)\rangle |\psi_i\rangle\langle\psi_j|=\sum_{i,j}\alpha_i\alpha^{\star}_{j}|\psi_i\rangle\langle\psi_j|,
\end{eqnarray}
where, as above, the sum is restricted to pairs $i, j$ such that $\lambda_i=\lambda_j$.

Since 
\begin{eqnarray}
\bar{P}_t(v|\alpha_0)=\sum_a\mathrm{tr}\left(|a,v\rangle\langle a,v|\frac{1}{t}\sum_{n=1}^{t}{\bf \Phi}_{\mathcal{A}}^n|\alpha_0\rangle\langle\alpha_0|\right),
\end{eqnarray}
and since
$\sum_a\mathrm{tr}\left(|a,v\rangle\langle a,v|\cdot\right)$ is a continuous function of the argument, it follows, by Theorem 7, that $\bar{P}_t(v|\alpha_0)$ converges to 
\begin{eqnarray}
\sum_{a}\mathrm{tr}\left(|a,v\rangle\langle a,v|\sum_{i,j}\alpha_i\alpha^{\star}_{j}|\psi_i\rangle\langle\psi_j|\right)=\sum_{i,j,a}a_ia^{\star}_j\langle a,v|\psi_i\rangle\langle \psi_j|a,v\rangle,
\end{eqnarray}
where the sum is restricted to pairs $i, j$ such that $\lambda_i=\lambda_j$. $\square$

\section{Classification of bistochastic quantum operations}

Let ${\bf \Phi}_{\mathcal{A}}:\mathfrak{B}(\mathcal{H})\rightarrow\mathfrak{B}(\mathcal{H})$ be a bistochastic quantum operation and let $\rho(0)\in \mathfrak{D}(\mathcal{H})$ denote the density matrix representing the initial state of a quantum system. Depending on the limiting behavior, as $t\rightarrow \infty$, of the corresponding quantum Markov process, ${\bf \Phi}_{\mathcal{A}}$ must belong to one of the following four mutually exclusive categories:

(1)\,\, $\lim_{t\rightarrow \infty}{\bf \Phi}_{\mathcal{A}}^t\rho(0)$ converges to the maximally-mixed state $\frac{1}{N}\mathbb{I}$, independently of the initial state $\rho(0)$.

(2)\,\, $\lim_{t\rightarrow \infty}{\bf \Phi}_{\mathcal{A}}^t\rho(0)$ converges, but the limit depends upon the initial state $\rho(0)$.

(3)\,\, $\lim_{t\rightarrow \infty}{\bf \Phi}_{\mathcal{A}}^t\rho(0)$ fails to converge, but the Ces$\grave{a}$ro average $\lim_{t\rightarrow \infty}\frac{1}{t}\sum_{n=1}^{t}{\bf \Phi}_{\mathcal{A}}^n\rho(0)$ exists and equals the maximally-mixed state $\frac{1}{N}\mathbb{I}$, independently of the initial state $\rho(0)$.

(4)\,\,  $\lim_{t\rightarrow \infty}{\bf \Phi}_{\mathcal{A}}^t\rho(0)$ fails to converge, but the Ces$\grave{a}$ro average $\lim_{t\rightarrow \infty}\frac{1}{t}\sum_{n=1}^{t}{\bf \Phi}_{\mathcal{A}}^n\rho(0)$ exists and depends upon the initial state $\rho(0)$.

To elucidate each of the above categories, we proceed to offer some comments and examples.

In category (1), the quantum operation ${\bf \Phi}_{\mathcal{A}}$ possesses no eigenvalue on the unit circle other than $\lambda =1$. Moreover, the eigenspace $\mathsf{Ker}({\bf \Phi}_{\mathcal{A}}-\mathbb{I})$ of $\lambda=1$ is one-dimensional, spanned only by the density operator $\frac{1}{N}\mathbb{I}$, or contains only a single density operator $\frac{1}{N}\mathbb{I}$. 

As an example, consider the quantum operation \cite{AA05} defined by 
\begin{eqnarray}
{\bf \Phi}_{\mathcal{A}}(X)=\frac{1}{2}X+ \frac{1}{2}UXU^{\dagger}
\end{eqnarray}
where $U$ is the unitary transformation given by 
\begin{eqnarray}
 U =
 \begin{bmatrix}
  \frac{1}{2} & -\frac{\sqrt{3}}{2}  \\
  \frac{\sqrt{3}}{2} & \frac{1}{2} 
 \end{bmatrix}.
\end{eqnarray}

According to \cite{NAJ10}, this is an example of a so-called {\em random unitary operation}. The eigenspace $\mathsf{Ker}({\bf \Phi}_{\mathcal{A}}-\lambda\mathbb{I})$ where $\lambda\in\Lambda_{1}({\bf \Phi}_{\mathcal{A}})$ is equal to the set $D_{\lambda}:=\{X\in \mathfrak{B}(\mathcal{H}): \mathbb{I}X=\lambda X\mathbb{I}\,\,\mbox{and}\,\,  UX=\lambda XU\}$. Evidently, $\Lambda_{1}({\bf \Phi}_{\mathcal{A}})=\{1\}$. A simple calculation shows that the eigenspace $\mathsf{Ker}({\bf \Phi}_{\mathcal{A}}-\mathbb{I})$ of the eigenvalue $\lambda=1$ is $D_{1}=\{k\mathbb{I}:k\in \mathbb{C}\}$, which is one-dimensional. Therefore $\lim_{t\rightarrow \infty}{\bf \Phi}^t_{\mathcal{A}}X=\frac{1}{2}\mathbb{I}_2$ for any initial state $X$. 

It can be verified that the {\em bit-phase flip} channel (p377 in \cite{NC00}) also belongs to category (1). A less trivial example of this sort can be found in \cite{LP10}.

In category (2), ${\bf \Phi}_{\mathcal{A}}$ possesses as its only eigenvalue on the unit circle the value $\lambda=1$ and $\mathsf{Ker}({\bf \Phi}_{\mathcal{A}}-\mathbb{I})$ is at least two-dimensional. 

As an example, consider the quantum operation ${\bf \Phi}_{\mathcal{A}}$ associated with a so-called {\em phase flip} channel on single qubits  \cite{NC00}, given by 
\begin{eqnarray}
{\bf \Phi}_{\mathcal{A}}(X)=pX+ (1-p)ZXZ^{\dagger},
\end{eqnarray}
where $Z$ is the {\em Pauli matrix}:  
\begin{eqnarray}
 Z =
 \begin{bmatrix}
  1 & 0  \\
  0 & -1 
 \end{bmatrix}.
\end{eqnarray}

By a pattern of reasoning similar to that employed in the previous example, we infer that $\Lambda_{1}({\bf \Phi}_{\mathcal{A}})=\{1\}$, and the eigenspace $\mathsf{Ker}({\bf \Phi}_{\mathcal{A}}-\mathbb{I})$ is spanned by the two linearly independent density operators:  
\begin{eqnarray}
 X_1 =
 \begin{bmatrix}
  1 & 0  \\
  0 & 0 
 \end{bmatrix},\,\,\,\,\,
X_2 =
 \begin{bmatrix}
  0 & 0  \\
  0 & 1 
 \end{bmatrix}.
\end{eqnarray}

In this case, $\lim_{t\rightarrow \infty}{\bf \Phi}^t_{\mathcal{A}}X=X_{\infty}$ where $X$ and $X_{\infty}$ are given by:

\begin{eqnarray}
 X =
 \begin{bmatrix}
  a_{11} & a_{12} \\
  a_{21} & a_{22} 
 \end{bmatrix},\,\,\,\,\,
X_{\infty} =
 \begin{bmatrix}
  a_{11} & 0  \\
  0 & a_{22} 
 \end{bmatrix}.
\end{eqnarray}
It can be verified without much difficulty that the {\em bit flip} channel (\cite{NC00}, p.376) belongs also to category (2). Additional examples of this sort can be found in (\cite{P08}, p.108). 

In category (3), the quantum operation ${\bf \Phi}_{\mathcal{A}}$ possesses at least two distinct eigenvalues (including $\lambda=1$) on the unit circle. The eigenspace of $\lambda=1$, namely $\mathsf{Ker}({\bf \Phi}_{\mathcal{A}}-\mathbb{I})$, is one-dimensional and spanned by $\mathbb{I}$. An example of this type of quantum operation is provided by {\em quantum walks} on the $N$-cycle. In this scenario, $N$ is assumed even, the Hadamard transform serves as the coin operator and the evolution of the system is subject to decoherence on both the position and the coin degrees freedom. A detailed treatment of this model is planned for our forthcoming paper \cite{LP11}.

In category (4), ${\bf \Phi}_{\mathcal{A}}$ has at least two distinct eigenvalues (including $\lambda=1$) on the  unit circle and $\mathsf{Ker}({\bf \Phi}_{\mathcal{A}}-\mathbb{I})$ contains at least two linearly independent density operators. As an example of this type of quantum operation, we cite \cite{NAJ10} which studies the properties so-called {\em two-qubit controlled-not} operators.\\

\section{Conclusion and related questions}
\noindent
Evidently, for a stochastic quantum operation ${\bf \Phi}_{\mathcal{A}}$, the eigenvalues lying on the unit circle determine the evolution of the associated quantum Markov process, including the existence or non-existence of a long-term stationary state. More precisely, the long-term behavior of the quantum Markov process is intimately linked to the structure of the eigenspaces of eigenvalues on the unit circle.  

We speculate that the eigenspaces of  eigenvalues on the unit circle, might conform always to a formulation in terms of a set of ``Kraus operators", and this formulation might provide an efficient means for identifying the eigenspaces of all eigenvalues of absolute 1. For {\it random unitary operations}, as in \cite{NAJ10}, and so-called {\it generalized random unitary operations}, as in \cite{LP11}, the structures of the aforementioned eigenspaces have been fully elaborated.

\section{Acknowledgment}

C. Liu was partially supported by NSF grant CCF-1005564.


\end{document}